\documentclass[12pt]{article}
\begin{document}
\begin{center}
 \textbf{DSR Relativistic Particle in a Lagrangian formulation and  Non-Commutative Spacetime: A Gauge Independent Analysis} \vskip 2cm

\textbf{Subir Ghosh}\\
Physics and Applied Mathematics Unit,\\
Indian Statistical Institute,\\
203 B. T. Road, Calcutta 700108, India.
\end{center}
\vskip 3cm
{\bf Abstract:}\\
In this paper we have constructed a coordinate space (or
geometric) Lagrangian  for a point particle that satisfies the
Doubly Special Relativity (DSR) dispersion relation in the
Magueijo-Smolin framework. At the same time, the symplectic
structure induces a Non-Commutative phase space, which
interpolates between $\kappa $-Minkowski  and Snyder phase space.
Hence this model bridges an existing gap between two conceptually
distinct ideas in a natural way.

We thoroughly discuss how this type of construction can be carried
out from a phase space (or first order) Lagrangian approach. The
inclusion of external physical interactions are also briefly
outlined.

   The work serves as a demonstration of how Hamiltonian (and
Lagrangian) dynamics can be built around a given non-trivial
symplectic structure.

\newpage
Quantum gravity motivates \cite{planck} us in accepting two
observer-independent scales: a length scale ($\sim$ Planck
length?) and of course, the velocity of light.  On the one hand,
this gave rise to the novel idea of Doubly Special Relativity
(DSR) \cite{am}. {\footnote{ One of the  motivations of studying
DSR is that it induces modified dispersion relations which might
be useful in explaining observations of ultra-high energy cosmic
ray particles, that violate the Greisen-Zatsepin-Kuzmin bound
\cite{am1}.}} On the other hand, existence of a length scale is
directly linked with the breakdown of spacetime continuum and the
emergence of a Non-Commutative (NC) spacetime (below {\it{e.g.}}
Planck length) \cite{sn}. The specific form of NC spacetime that
one supposedly encounters in the context of DSR is known as $\kappa
$-Minkowski spacetime \cite{dsr2}. There are manifestly distinct
constructions of DSR dispersion relation
\cite{dsr1,dsr2,dsr3,mag,kow}, which appear as Casimir operators
of distinct $\kappa $-Poincare algebra. From Quantum Group
theoretic point of view, it has been argued \cite{kow,kow1} that
these $\kappa $-Poincare algebra are {\it{dual}} to different
forms of $\kappa $-Minkowski NC phase space, all of which, indeed,
have the {\it{same}} $\kappa $-Minkowski spacetime structure.
Indeed, this state of affairs makes the connection between DSR and
NC spacetime somewhat indirect.

In the present state of affairs, as described above, there exist two important gaps that make the physical picture of DSR particle incomplete. These are the following:\\
 (i) A {\it{minimal}}
$3+1$-dimensional {\it{dynamical}} model that induces the
Magueijo-Smolin  (MS) DSR dispersion relation is absent.\\
 (ii) A
{\it{direct connection}} in a dynamical framework between the DSR particle  and its
underlying NC spacetime is lacking.

The present work sheds light on both these areas simultaneously.
As for (i), we provide an explicit geometric Lagrangian that obeys
the exact DSR dispersion law in Magueijo-Smolin base. The problem
(ii) is also taken care of since the particle phase space discussed here has an
inherent NC structure, which turns out to be a novel one. It is a
combination between $\kappa $-Minkowski (in MS base) and Snyder
\cite{sn} NC phase space algebra. This algebra emerges as Dirac
Brackets after an analysis of the constraints are performed in
Hamiltonian framework \cite{dirac}. Note that our conclusion deviates
somewhat from the previously obtained Quantum Group motivated
result \cite{dsr2,dsr1,dsr3,kow}
of an association between the MS particle and $\kappa $-Minkowski
phase space (in MS base). The latter derivation does not involve particle
dynamics and is more mathematically inclined.

The present work is also an improvement from our previous attempt
\cite{sgprd} where the  NC phase space was recovered in a
particular gauge.

The plan of the paper is the following: In Section I we will
present the geometric Lagrangian of the DSR particle in
Magueijo-Smolin base and demonstrate that it correctly reproduces
the MS DSR relation. We also perform the constraint analysis to
reproduce the interpolating NC phase space algebra. This is the
main result of our work. In Section II we study the symmetry
properties of this NC phase space and show that DSR mass-shell
condition, reproduced here, is Lorentz invariant. Section III is
devoted to a systematic analysis of the derivation of this
particular geometric Lagrangian from a first order phase space
Lagrangian \cite{rb}. In Section IV we summarize our results and
point out some future directions of study.

\vskip .5cm {\bf{Section I: The DSR Geometric Lagrangian}} \vskip
.2cm
The point particle dispersion relation in Magueijo-Smolin base \cite{mag} is,
\begin{equation}
p^2=m^2(1-\frac{E}{\kappa })^2,  \label{0m}
\end{equation}
where $E$ denotes the particle energy $E=p^0$. We posit the following geometric Lagrangian in order to describe
the point particle obeying (\ref{0m}),
\begin{equation}
L=\frac{\kappa }{D}\frac{x^2}{(\eta x)^2}[\sqrt B +\frac{(\eta
 x)(x \dot x) }{x^2}-(\eta \dot x)].
 \label{n7}
\end{equation}
In the above the functions $B$ and $D$ stand for,
$$
B=\{((\eta \dot x)-\frac{(x\dot x)(\eta x)}{x^2})^2+(\frac{(\eta
x)^2 }{x^2}+\frac{\kappa ^2-m^2}{m^2})(\dot x^2-\frac{(x\dot x)^2
}{x^2})\},$$
\begin{equation}
D=1+\frac{(\kappa ^2-m^2)}{m^2}\frac{x^2}{(\eta x)^2}.
 \label{nn6}
\end{equation}
Our metric is $diag~g^{00}=-g^{ii}=1$. The constant vector $\eta _\mu \equiv \eta _0=1,\eta _i=0$ is introduced to maintain relativistic notation. Clearly the presence of $E=\eta^\mu p_\mu $ in (\ref{0m}) makes the relation non-covariant from Special Theory of Relativity point of view. $\eta_{\mu}$ will also be useful later when we discuss the NC phase space algebra. The notation $a^\mu b_\mu =(ab)$ is followed throughout. The Lagrangian in (\ref{n7}) constitutes our principal
result.

First we reproduce the DSR dispersion relation.  The conjugate
momentum
$$p^\mu =\partial L/\partial
\dot x_\mu $$ is computed as,
$$
p_\mu =\frac{\kappa x^2}{D(\eta x)^2}[\{\eta _\mu -\frac{(\eta
x)}{x^2}x_\mu \}+\frac{1}{\sqrt B}\{(\eta \dot x)-\frac{(x\dot x)
(\eta x)}{x^2}\}\{\eta _\mu -\frac{(\eta x)}{x^2}x_\mu
\}$$
\begin{equation}
+D\frac{(\eta x)^2}{x^2}\{\dot x_\mu -\frac{(x\dot
x)}{x^2}x_\mu\}]
 \label{n8}
\end{equation}
It is straightforward but tedious to check that (\ref{n8})
satisfies the relation,
\begin{equation}
p^2=m^2(1-\frac{(\eta p)}{\kappa })^2 =m^2(1-\frac{E}{\kappa })^2.  \label{m}
\end{equation}

Next we recover the NC  phase space structure. This is achieved
through a Hamiltonian analysis of constraints, as formulated by
Dirac \cite{dirac}. Notice that apart from the MS mass-shell
condition (\ref{m}), there is another primary constraint, that
follows from (\ref{n8}):
\begin{equation}
\psi _2\equiv (xp)\approx 0. \label{nn8}
\end{equation}
For this set of constraints
\begin{equation}
\psi _1\equiv p^2-m^2(1-\frac{(\eta
p)}{\kappa })^2\approx 0~;~~
\psi _2\equiv (xp)\approx 0,
\label{cons}
\end{equation}
 we find that their Poisson Bracket is  non-vanishing,
\begin{equation}
 \{\psi _1
,\psi _2 \}=2m^2(1-(\eta p)/\kappa ). \label{01}
\end{equation}
In the terminology of Dirac,  non-commutating and  commutating
 (in the sense of Poisson Brackets) constraints are termed as
Second Class Constraints (SCC)  and First Class Constraints (FCC)
respectively. Presence of SCCs require a modification in the
symplectic structure by way of replacing Poisson Brackets by Dirac
Brackets. On the other hand, FCCs  induce local gauge invariance.

In the presence of  SCCs $\psi_{i}$ with $\{\psi_1 ,\psi _2 \}\ne
0$, the  Dirac Brackets are defined in the following way,
\begin{equation}
\{A,B\}^*=\{A,B\}-\{A,\psi _i\}\{\psi _i,\psi _j\}^{-1}\{\psi
_j,B\}, \label{n4}
\end{equation}
where $\{\psi _i,\psi _j\}$ refers to the invertible  constraint
matrix. From now on we will always use Dirac brackets and refer
them simply as $\{A,B\}$.

For the model at hand, we find the only non-vanishing constraint
matrix component to be
$$
\{\psi_1 ,\psi _2 \}=2m^2(1-(\eta p)/\kappa ). $$
 From (\ref{n4})  the Dirac brackets
follow:
$$
\{x_\mu ,x_\nu \}=\frac{1}{\kappa}(x_\mu \eta_{\nu}-x_\nu
\eta_{\mu })+\frac{1}{m^2(1-(\eta p)/\kappa)}(x_\mu p_{\nu}-x_\nu
p_{\mu }),$$
\begin{equation}
\{x_{\mu},p_{\nu}\}=-g_{\mu\nu}+\frac{1}{\kappa}\eta_{\mu}p_{\nu}+\frac{p_{\mu}p_{\nu}}{m^2(1-(\eta
p)/\kappa)},~~\{p_{\mu},p_{\nu}\}=0.
 \label{03}
\end{equation}
Performing an (invertible) transformation on the variables,
\begin{equation}
\tilde x_\mu =x_\mu -\frac{1}{\kappa }(x\eta )p_\mu ,~~x_\mu =\tilde x_\mu +\frac{(\eta \tilde x)}{\kappa (1-(\eta p)/\kappa )}p_\mu ,
 \label{04}
\end{equation}
we find  a novel and interesting form of algebra that interpolates
between the Snyder \cite{sn} and $\kappa $-Minkowski form
\cite{am,dsr2,kow}:
$$
\{\tilde x_\mu , \tilde x_\nu \}=\frac{1}{\kappa }(\tilde x_\mu
\eta _\nu -\tilde x_\nu \eta _\mu )+\frac{\kappa ^2 -m^2}{\kappa
^2 m^2}(\tilde x_\mu p_\nu -\tilde x_\nu p _\mu ),$$
\begin{equation}
\{\tilde x_\mu , p_\nu \}= -g_{\mu\nu}+\frac{1}{\kappa }(p_\mu
\eta _\nu +p_\nu \eta _\mu )+\frac{\kappa ^2 -m^2}{\kappa ^2 m^2}p
_\mu p_\nu ~,~\{p_\mu , p_\nu \}=0. \label{05}
\end{equation}
In absence of the $1/\kappa $-term  one obtains the Snyder
\cite{sn} algebra,
$$
\{\tilde x_\mu , \tilde x_\nu \}=\frac{\kappa ^2 -m^2}{\kappa ^2
m^2}(\tilde x_\mu p_\nu -\tilde x_\nu p _\mu ),$$
\begin{equation}
\{\tilde x_\mu , p_\nu \}= -g_{\mu\nu}+\frac{\kappa ^2
-m^2}{\kappa ^2 m^2}p _\mu p_\nu ~,~\{p_\mu , p_\nu \}=0.
\label{sn}
\end{equation}
On the other hand, without the $(\kappa ^2 -m^2)/(\kappa ^2 m^2)$ term, one finds the  $\kappa
$-Minkowski phase space in MS base  \cite{kow,kow2},
$$
\{x^\mu ,x^\nu \}=\frac{1}{\kappa} (x^\mu \eta ^\nu -x^{\nu}\eta^{\mu} ),
$$
\begin{equation}
\{x^\mu ,p^\nu \}=-g^{\mu\nu}+\frac{1}{\kappa} (p^\mu \eta ^\nu +p^\nu
\eta^{\mu} )~; \{p^\mu ,p^\nu \}=0. \label{kap}
\end{equation}
It is important to mention the base in which the $\kappa
$-Minkowski algebra is expressed (see for example
\cite{kow,kow2}). The algebra (\ref{kap}) in detailed form is,
$$
\{x^i,x^j\}=0~;~~\{x^0,x^i\}=-\frac{1}{\kappa} x^i, $$
$$
 \{x^0,p^i\}=\frac{1}{\kappa}
p^i~;~\{x^i,p^0\}=\frac{1}{\kappa} p^i~;~\{x^0,p^0\}=-1+\frac{2}{\kappa} p^0 ,$$
\begin{equation}
\{x^i,p^j\}=-g^{ij}~;~\{p^\mu,p^\nu\}=0. \label{2}
\end{equation}
 For $\kappa =\infty $ one recovers
the normal canonical phase space. Note that choice of other forms
of  bases, (such as bi-crossproduct or standard base), leads to
forms of $\{x_\mu,p_\nu \}$ algebra that are distinct from
(\ref{2}) but in {\it{all}} the bases, the spacetime algebra is
identical to that of (\ref{2}) \cite{desit}.

As we have mentioned in the Introduction, the NC phase space that we have obtained, turns out to be a generalized version of $\kappa $-Minkowski algebra and in fact reduces to the latter for the choice $m=\kappa $. This choice of $m$, the rest mass, is the upper limit for which the MS relation becomes  non-relativistic, $p_0=(\vec p)^2/(2\kappa )+\kappa /2$.

\vskip .5cm {\bf{Section II: Symmetry Properties }} \vskip .2cm It
is well known that {\it{individually}} both  Snyder and $\kappa
$-Minkowski  extensions   of the Poincare algebra, keep the
Lorentz subalgebra sector  unchanged and only transformation
properties under boost are modified \cite{desit}. It is reassuring
to find that this feature holds for the present interpolating
algebra (\ref{05}) as well, which happens to be a combination of
Snyder and $\kappa $-Minkowski  algebra.

 Let us  consider the conventional
form of Lorentz generators
$L^{\mu\nu}=x^{\mu}p^{\nu}-x^{\nu}p^{\mu}$ and using (\ref{05}) we find that the
Lorentz algebra remains intact,
\begin{equation}
\{L^{\mu\nu },L^{\alpha\beta }\}=g^{\mu\beta }L^{\nu\alpha
}+g^{\mu\alpha }L^{\beta \nu}+g^{\nu\beta }L^{\alpha\mu
}+g^{\nu\alpha }L^{\mu\beta }. \label{nn2}
\end{equation}
However,  $x^{\mu}$ and $p^{\mu}$ transform in the following
non-trivial way:
\begin{equation}
\{L^{\mu\nu},x^\rho \}=g^{\nu\rho}x^\mu-g^{\mu\rho}x^\nu
-\frac{1}{\kappa} (\eta^\mu L^{\rho\nu}-\eta^\nu L^{\rho\mu })~;~
\{L^{\mu\nu},p^\rho \}=g^{\nu\rho}p^\mu -g^{\mu\rho}p^\nu
-\frac{1}{\kappa} (\eta^\nu p^\mu-\eta^\mu p^\nu)p^\rho
.\label{a2}
\end{equation}
Notice that only the boost relations are modified \cite{mag} (see also
Kowalski-Glikman in \cite{am}),
\begin{equation}
\{L^{0i},x^\mu \}= x^0g^{i\mu}-x^ig^{0\mu}-\frac{1}{\kappa} L^{\mu
i}~;~ \{L^{0i},p^\mu \}= p^0g^{i\mu}-p^ig^{0\mu}+\frac{1}{\kappa}
p^ip^\mu .\label{a6}
\end{equation}

We note that the mass-shell condition (\ref{m}) is Lorentz
 invariant on-shell:
 \begin{equation}
\{L^{\mu\nu},p^2-m^2(1-\frac{(\eta p)}{\kappa
})^2\}=-\frac{2}{\kappa }(p^\mu \eta ^\nu - p^\nu \eta ^\mu
)(p^2-m^2(1-\frac{(\eta p)}{\kappa })^2). \label{nn1}
\end{equation}
\vskip .5cm {\bf{Section III: Coordinate space Lagrangian }}
\vskip .2cm Our objective is now to recover the coordinate space
geometric Lagrangian  (\ref{n7}), stated at the beginning. We follow the procedure
described in a recent work \cite{rb}. Since the relativistic
particle model  is reparameterization invariant, it has a
vanishing Hamiltonian and the action will consist of the normal
kinetic term and constraints:
\begin{equation}
S=\int d\tau [(p\dot x)-\frac{\lambda
_1}{2}\{p^2-m^2(1-\frac{(\eta p)}{\kappa })^2\}-\lambda _2(xp)]\equiv \int d\tau [(p\dot x)-\frac{\lambda
_1}{2}\psi _1 -\lambda _2\psi _2 ].
\label{n1}
\end{equation}
The idea is to exploit the equation of motion along with the
constraint conditions to express the multipliers $\lambda _1$ and
$\lambda _2$ and $p_\mu $ in terms of coordinate ($x_\mu $) and
velocity ($\dot x_\mu $) variables. The equation of motion is,
\begin{equation}
\dot x_\mu -\lambda _1\{p_\mu +\frac{m^2}{\kappa }(1-\frac{\eta
p)}{\kappa })\eta _\mu \} -\lambda _2 x_\mu =0.
 \label{n2}
\end{equation}
The above is rewritten as,
\begin{equation}
p_\mu =\frac{1}{\lambda _1}[\dot x_\mu -\lambda _2x_\mu
+\frac{m^2}{\kappa ^2 -m^2}\{(\eta \dot x)-\kappa \lambda _1-(\eta
x)\lambda _2\}\eta _\mu ].
 \label{n3}
\end{equation}
The constraint $(xp)=0$ yields a relation between $\lambda _1$ and
$\lambda _2 $. Subsequently we substitute this relation together
with (\ref{n3}) in the MS mass-shell constraint relation $\psi _1=0$ and obtain
an algebraic  quadratic equation for $\lambda _2$. For $\lambda _2$ we take the solution with the positive root of the discriminant and then in
turn we compute $\lambda _1$. The expressions are,
\begin{equation}
\lambda _2=\frac{1}{D}[\{\frac{(\eta \dot x)}{(\eta
x)}+\frac{\kappa ^2 -m^2}{m^2}\frac{(x \dot x)}{(\eta x)^2}\}+
\frac{1}{(\eta x)}\sqrt B],
 \label{n5}
\end{equation}
\begin{equation}
\lambda _1=\frac{1}{\kappa }\sqrt B .
 \label{n6}
\end{equation}

 Finally we are able to get  $p_\mu $ in terms of
$x_\mu $ and $\dot x_\mu $ only by using (\ref{n3}). This yields the
cherished expression for the Lagrangian that was provided at the
beginning in (\ref{n7}).

\vskip .5cm {\bf{Section IV: Summary and Future Outlook }} \vskip
.2cm In the present work, for the first time a fully dynamical
model of a   particle having the (Magueijo-Smolin)  DSR dispersion
relation has been provided in the {\it{Lagrangian framework}}. The
connection between DSR and NC phase space is also attained in a
direct way since the model enjoys an inherent NC phase space. The
NC algebra is in itself novel and interesting since it
interpolates between the normal Snyder algebra and $\kappa
$-Minkowski algebra. There is also an improvement (in an esthetic
way) compared to previous works \cite{g,sgprd} since {\it{no gauge fixing is
necessary}}.

We list some of the problems that we intend to study in near
future:\\
(i) It is very important to study the  behavior of the DSR
particle in presence of interaction. So far this has not been done
mainly because a suitable Lagrangian formulation of the model was
lacking. Indeed, in our approach, an external interaction can be
easily included, the simplest being a minimal coupling to a $U(1)$
gauge field and we end up with the interacting system,
\begin{equation}
L_{I}=\dot x_\mu  p^{\mu}+{\cal{A}}_{\mu}(x)\dot x^\mu
-\frac{\lambda _1}{2}\psi _1 - \lambda _2\psi _2, \label{int}
\end{equation}
where ${\cal{A}}_{\mu}(x)$ is the external {\it{physical}} gauge
field.  \\
(ii) It will be interesting to construct, in the present framework,
point particle models of other  DSR structures. It has been shown
that these different DSR bases are related through non-linear
transformations of variables and explicit rules are provided in
\cite{desit}. However, whether this (mathematical) equivalence
generates physically equivalent models can be tested by
considering dynamical models of the present form, where the
particle behavior can be studied in a direct way.
 \vskip .5cm
 {\it{Acknowledgement}}: It is a pleasure to thank  Professor J. Lukierski for
 discussions.

\end{document}